\documentstyle[aps,preprint,prl]{revtex}
\def\beq{\begin{equation}}
\def\eeq{\end{equation}}
\begin{document}                
\title{On the theory of the spin gap in bilayer cuprates}
\author{L. B. Ioffe\cite{landau}}
\address{Serin Physics Laboratory, Rutgers University}
\author{A. I. Larkin\cite{landau}, A. J. Millis}
\address{A. T. \& T. Bell Labs, 600 Mountain Ave. Murray Hill, NJ 07974}
\author{B. L. Altshuler}
\address{Physics Department, MIT, Cambridge, MA 02139}
\maketitle
\begin{abstract}
We formulate and solve a model of two planes of antiferromagnetically
correlated electrons coupled together by a weak antiferromagnetic interaction
of strength $\lambda$.
We show that in-plane antiferromagnetic correlations dramatically enhance
the pairing effect of the interplane interaction.
For the case where the in-plane correlation length $\kappa^{-1} \sim T^{-1/2}$,
we find that the interaction $\lambda$ leads to spin pairing at a temperature
$T^{*} \sim \lambda$, much higher than the usual BCS result $exp(-J/\lambda)$.
We suggest that this is a possible explanation of the spin gap effects
observed below $T^{*} \sim 150K$ in $YBa_2Cu_3O_{6.6}$.
\end{abstract}

\narrowtext

It was recently argued that superconductivity and spin gaps in bilayer
copper oxides such as $YBa_2Cu_3O_{6+x}$ may be due to interplane pairing
\cite{altshuler,millis1,geshkenbein} caused by the antiferromagnetic
spin-spin interaction between the planes.
Effects of this interaction have been observed in neutron scattering
experiments on  $YBa_2Cu_3O_{6.6}$ \cite{neutrons}.
The high-$T_c$ materials are also believed to have strong in-plane
antiferromagnetic fluctuations.  An alternative mechanism for spin gap
formation in copper oxide materials, based on a single-plane theory
of bosonic spin-waves, has also been discussed \cite{Sachdev}.
In this paper we determine the effect of the in-plane fluctuations on the
interplane pairing interaction discussed previously.
We find that they strongly enhance the between-planes interaction at wave
vectors near the wave vector $\bf Q$ where the in-plane spin susceptibility is
peaked.
Taking into account this enhancement and the modification of the electron
spectrum by the spin fluctuations \cite{millis2} we obtain an estimate for the
onset temperature for the spin gap which is of the correct order of magnitude.

Several different cases arise which we discuss in detail elsewhere
\cite{longpaper}.
One issue is the relation between vector $\bf Q$ and the shape of the Fermi
surface of the fermions: the vector $Q$ might be a chord of the Fermi surface,
its diameter and be larger than $2p_F$.
\footnote{all these conditions have trivial generalization for a non-spherical
Fermi surface, for brevity we discuss only the circular case here.}
In this paper we consider only the chord case.
Another issue concerns the strength of the spin correlations.
Here we assume that the spin system in each plane is very close to a $T=0$
critical point \cite{criticality} resulting in long range antiferromagnetic
fluctuations with a correlation length proportional to a power of temperature.
A third issue is the nature of the fermionic excitations.
One may distinguish the "spin liquid" case with spin-charge separation and
fermionic spin excitations \cite{anderson} and the
"Fermi liquid" case, where there is no spin-charge separation.
Formally, the difference between these two pictures originates from the
presence of an additional low energy mode ( gauge field ) in the spin liquid
case \cite{gauge},
which results in a large relaxation rate for the fermions (so the
electron propagator is $(\beta \epsilon^{2/3} -v_F|p-p_F|)^{-1}$ \cite{lee1}).
We treat both cases here.
It is also believed that in underdoped high-$T_c$ materials electrons cannot
tunnel coherently between planes \cite{rperp}, so we shall assume that
all low energy excitations are confined to a plane.

To model one plane of antiferromagnetically correlated fermions we write
\beq
H_{F}=\sum_{p}  c^{\dagger}_{p\sigma} \epsilon(p) c_{p\sigma} +
\sum_{q} J_q {\bf S}_{q} {\bf S}_{-q}
\label{H_F}
\eeq
where $\epsilon(p)=v_F(|p|-p_F)$ is the fermion dispersion near the Fermi
surface,
${\bf S}_{i}=c^{\dagger}_{i\sigma} {\bf \sigma}_{\alpha\beta} c_{p\sigma}$.
It is also convenient to introduce the fermion mass $m=p_F/v_F$.
The interaction $J_q$ causes antiferromagnetic correlations peaked at wave
vector {\bf Q}.
For definiteness we treat the interaction in the RPA approximation and assume
that the parameters are such that the spin susceptibility $\chi(k,\omega)$ is
given by
\begin{eqnarray}
\chi(k,\omega)&=& \frac{\chi_0(k,\omega)}{1-J_k\chi_0(k,\omega)} \\
          &=& \frac{J_Q^{-1}}{\kappa^2 + ({\bf k-Q})^2 + |\omega|/\Gamma }
\label{chi}
\end{eqnarray}
where $\chi_0$ is the susceptibility of the non-interacting fermions, $\kappa$,
the inverse correlation length, is assumed to be small and $\Gamma$ is a
'microscopic' frequency scale. Presumably $\Gamma \sim 1/m$, or $\Gamma \sim
J/p_F^2$.
To fit the $Cu$ NMR relaxation rates in high $T_c$ materials at high
temperatures it is necessary to take $\kappa^2=MT$ where $M$ is a constant.
We emphasize that although we have used the RPA to explain the form of
(\ref{chi}), this form is more general than the explanation \cite{millis2}
and so are the following results which depend on (\ref{chi}) only.
The specific form of (\ref{chi}) holds only if the wave vector $Q < 2p_F$, so
that at all wave vectors near $\bf Q$ a particle hole pair is available to
damp the spin excitation.

In the following we choose polar coordinates on
the Fermi surface so that the points on the Fermi surface connected by
$\bf Q$ correspond to angles $\pm \theta_0$.
The form (\ref{H_F}) applies to both the 'spin liquid' and Fermi liquid cases.
In $Q=2p_F$ case the functional form of $\chi$ is different and depends on
whether the fermion damping is small or large.

We assume that the only coupling between different planes is an
antiferromagnetic interaction between spins:
\beq
H_{int} = \lambda \sum_i {\bf S}_i^{(1)} {\bf S}_i^{(2)},
\label{H_int}
\eeq
where indices 1 and 2 distinguish planes in a bilayer and $\lambda$ is an
interaction constant, assumed small.
Neutron measurements \cite{neutrons} imply that $\lambda \sim 200K$, but
certainly $\lambda
\ll J$ where $J\sim 1500K$ is the exchange constant in one plane.

The interaction (\ref{H_int}) leads to antiferromagnetic correlations between
planes which we assume to be weaker than the in-plane correlations.
An arbitrary weak $\lambda$ has also been shown \cite{altshuler} to lead to a
singlet pairing of spin excitations in different planes.
In this work the antiferromagnetic correlations within each plane were not
taken into and the temperature at which the spin pairing occurred was found to
be very low ($T_c \sim \epsilon_F e^{-\lambda/\epsilon_F}$).
Here we show that in the presence of antiferromagnetic correlations the
pairing interaction becomes much stronger at wave vectors near $\bf Q$, the
temperature at which the pairing occurs is greatly enhanced and the gap
function becomes very anisotropic,
opening first in a small region ( about $(\theta-\theta_0) \sim\kappa/p_F$ )
around the
points connected by the vector $\bf Q$ and dropping rapidly away from these
points as $1/(\theta-\theta_0)^4$.

The physical argument is that because the susceptibility in one plane is very
large at wave vectors near $\bf Q$, a fermion at this wavevector polarizes the
electrons in the neighboring plane in a large area around itself.
Mathematically, we must construct the pairing vertex connecting a particle in
one plane to a particle in the other.
For small $\lambda$ this vertex will be linear in $\lambda$ and will be dressed
by spin fluctuations in each plane: within RPA we have found that the dominant
contribution to the dressed vertex $V(k,\omega)$ is that shown on fig 1 which
leads to

\beq
\begin{array}{ll}
H_{int}^{d} = \sum_{p,p',k} V(k,\omega) c_{p+k}^{\dagger}\sigma^{\alpha}c_{p}
                c_{p'+k}^{\dagger} \sigma^{\alpha} c_{p'} \nonumber \\
V(k, \omega)= \lambda J_Q^2 a^{-2}\chi^2(k, \omega)
\end{array}
\label{H_int^d}
\eeq
where $a$ is the lattice constant.
Other contributions are negligible.
To calculate the onset of the pairing from Eq. (\ref{H_int^d}) we must sum the
ladder diagrams shown in Fig. 2.
It is important to use the full Green's function, including the self energy
due to spin exchange within one plane.
This self energy has been studied by many authors.  An approximation convenient
for our purposes is \cite{millis3}:
\beq
\Sigma(\omega,\theta) = \frac{\alpha_Q |\omega| J m}
		{2\pi p_F\sqrt{\omega/\Gamma +p_F^2 (\theta-\theta_0)^2 +
	\kappa^2}}
\label{Sigma}
\eeq
where $\alpha_Q$ is a function of the order of unity if $\bf Q$ is away from
$2 p_F$ but which diverges as $Q \rightarrow 2p_F$.  We have verified that
this formula applies also in the spin-liquid case.

It is important that the gap is due to the pairing of spin excitations on
different
planes, so the interaction (\ref{H_int^d}), although large in some region of
momentum space, does not lead  to  self energy parts or vertex
corrections.
The gap equation, thus, follows from the summation of the ladder series in Fig.
2.  By performing the ladder sum and  integrating over the momenta in the
direction normal to the Fermi  surface we get
\beq
\Delta(\epsilon,\theta)=\frac{T}{4\pi}\sum_{\omega} \int
\frac{\lambda m \Delta(\epsilon+\omega,\theta') d\theta' }
{[|\omega|/\Gamma+ p_F^2(\theta^2+\theta'^2 +2u\theta\theta') +\kappa^2]^2
\sqrt{[\omega+\Sigma(\omega)]^2+\Delta(\epsilon+\omega,{\theta'})^2}}
\label{Delta}
\eeq
where $u=1-Q^2/(2p_F)^2$ and we have set $\theta_0 =0$.
The integration over the perpendicular momenta was possible because the main
contribution to this integral comes from a narrow range near the Fermi surface
($\delta p' \sim T/v_F$) where the interaction $V({\bf p-p'},\omega)$ does not
vary significantly.

To find the onset temperature we linearize (\ref{Delta}) and introduce
scaled variables x and y via $\theta=\kappa x/p_F$ and $\theta'=\kappa y/p_F$.
The resulting equation is
\beq
\Delta_n(x)=\frac{\lambda}{2 \alpha_Q  MT a^2 J}
    \sum_{l} \int dy
    \frac{\Delta_{n+l}(y)\sqrt{1+y^2+\frac{2\pi}{M\Gamma}|n+l+\frac{1}{2}|}}
		{|n+l+\frac{1}{2}|(1+y^2+x^2+2uxy+\frac{2\pi}{M\Gamma}|l|)^2}
\label{Delta_n}
\eeq
where $l$ and $n$ are integers.
{}From (\ref{Delta_n}) it is evident that $\Delta$ depends only on $1+x^2$
which
enters only in the denominator of the kernel.
Thus, $\Delta(\theta)$ is peaked about $\theta=0$ with a width $\kappa$ and
decays for large $\theta$ as $1/\theta^4$, and is similarly peaked about the
lowest Matsubara frequency $\omega_n=\pi T$ with the width $\Gamma \kappa^2
\sim T$.
The dimensionless kernel in (\ref{Delta_n}) presumably has a largest
eigenvalue $w \sim 1$, so $T^{*}$ is given by
\beq
T^{*} = \frac{w \lambda}{2 \alpha_Q M a^2 J}
\label{T^*}
\eeq
Thus, apart from numerical factors the onset temperature $T^*$ is given by the
bare interplane coupling constant $\lambda$.
For $T \ll T^*$ we may replace the sum over frequencies in (\ref{Delta})
by an integral; this integral is dominated by frequencies of the order of the
zero temperature spin gap $\Delta(0)=\Delta^*$; similarly, we must replace
$\kappa^2$ by $M \Delta^*$ because low frequency spin correlations near the
antiferromagnetic wavevector $\bf Q$ are eliminated by the spin gap.
The result is that up to numerical factors $T^*$ in (\ref{T^*}) is replaced by
$\Delta^*$.
The gap takes its maximum value for angles $\theta \lesssim \sqrt{M\Delta^*}$;
for larger $\theta$ it given by $\Delta(\theta)\sim\Delta^*
\left[\frac{\kappa}{p_F \theta} \right]^4 \sim
\frac{(\Delta^{*})^3 M^2}{p_F^2 \theta^4} $.

We emphasize that due to a strongly peaked and temperature dependent effective
interaction, the pairing
temperature and the gap scale as the interaction constant, unlike the
usual BCS case where they are exponentially small.
Although we have assumed specific form for the spin susceptibility (\ref{chi})
with a temperature dependent correlation length, this assumption is not
essential to our results.
The enhancement of the between-planes pairing is due to the strong temperature
dependence of $\sum_q \chi'(q,\omega=0)^2$.
This quantity is measurable via NMR $T_2$ experiments \cite{pemington} and has
been found to be large and strongly temperature dependent in
$YBa_2Cu_3O_{7-x}$ \cite{pemington,machi}.

The development in the spin liquid case is essentially identical.
The momentum integrated Green's function square is $[\beta |\omega|^{2/3} -
\Sigma(\omega ,\theta)]^{-1}$ but $\beta |\omega|^{2/3}$ is still negligible
compared
to $\Sigma(\omega ,\theta)$ for the frequencies and angles of interest, so Eqs
(\ref{Delta_n}-\ref{T^*}) are not changed.

We now consider the experimental implications.  The pairing mechanism is much
weaker in $La_{2-x}Sr_xCuO_4$ because the antiferromagnetic interaction between
Cu ions in different planes is frustrated, so that in tetragonal crystals Eq
(\ref{H_int}) becomes
\begin{equation}
H_{int}^{La}= \lambda \sum_{i,\delta} {\bf S}_i^{(1)} {\bf S}_{i+\delta}^{(2)}
\label{H_int^La}
\end{equation}
where $\delta$ labels the four Cu sites in plane 2 equidistant from site i
of plane 1.  Eq (\ref{H_int^La}) implies that $V(k,\omega)$ in eq.
(\ref{H_int^d}) becomes
\begin{equation}
V^{La}(k,\omega) = V(k,\omega) \cos(k_x/2) cos(k_y/2)
\label{H_int^d^La}
\end{equation}
Thus the singularity in the interaction is eliminated for commensurate spin
fluctuations ($k_x,k_y \sim \pi$) in tetragonal crystals.  For orthorhombic
crystals or for incommensurate spin fluctuations the singular part of the
interaction is of order the square of the orthorhombicity or
incommensurability,
and is therefore small.  This is consistent with the observation that the
spin gap opens at much lower temperatures in $La_{2-x}Sr_xCuO_4$ than in
$YBa_2Cu_3O_{6+x}$.

In a Fermi liquid system with no spin-charge separation the opening of the
spin gap implies that the material has become superconducting.  In a spin
liquid system, true superconductivity will only occur at a lower temperature
where the charge carriers bose condense.
The former scenario is consistent with the behavior
of optimally doped or overdoped $YBa_2Cu_3O_{6+x}$ and with $La_{2-x}Sr_xCuO_4$
at all dopings, while the latter scenario is consistent with the behavior
of underdoped $YBa_2Cu_3O_{6+x}$.  For example, in  $YBa_2Cu_3O_{6.6}$, spin
gap effects are observed in NMR below $T^* \sim 150K$ while the superconducting
$T_c \sim 60K$.
As previously pointed out \cite{altshuler} there is also
optical evidence \cite{Thomas} for the existence of a gap above $T_c$.
The small value of the specific heat jump at $T_c$ in $YBa_2Cu_3O_{6.6}$
\cite{loram} is also consistent with this scenario.  However, none of
these observations (except the qualitative one that the spin gap
opens significantly above $T_c$ only in underdoped bilayer materials such
as $YBa_2Cu_3O_{6.6}$) distinguishes the mechanism we have proposed from
other possible origins of the spin gap.

There is one qualitative disagreement with experiment.  Because the gap opens
first and is largest at the points on the Fermi surface connected by the
wavevector where $\chi(k,\omega)$ is peaked, the low frequency
antiferromagnetic
spin fluctuations are suppressed more strongly than spin fluctuations at
other wavevectors.  In the high-$T_c$ materials it is believed that the
antiferromagnetic fluctuations are responsible for the enhancement
of the Cu relaxation rate over the relaxation rates of the other nuclei
\cite{criticality}; therefore in our scenario the copper relaxation rate
would
drop more rapidly than the oxygen or yttrium rates as the spin gap opened,
in apparent disagreement with experimental data on $YBa_2Cu_3O_{6.6}$
\cite{Takigawa}.

Note added:  As this manuscript was being prepared we learned that M. Ubbens
and P. A. Lee \cite{lee2} have obtained results very similar to ours.

\begin{figure}
\caption{Diagrammatic representation of dominant contribution to pairing
interaction. Here the light dashed line represents the interplane interaction
and the wavy line represents the dressed spin-fluctuation interaction between
electrons in one plane.}
\end{figure}
\begin{figure}
\caption{Ladder sum leading to gap equation. Here the shaded
rectangle is the interaction $V$ defined in Fig 1. Note that electron lines
are dressed by in-plane fluctuations.}
\end{figure}

\end{document}